\documentclass[twocolumn,showpacs,preprintnumbers,amsmath,amssymb,nofootinbib]{revtex4}
\usepackage{graphics,epsfig,subfigure}
\usepackage{color}
\usepackage[linktocpage]{hyperref}

\begin{document}
\renewcommand{\baselinestretch}{1.3}
\newcommand\beq{\begin{equation}}
\newcommand\eeq{\end{equation}}
\newcommand\beqn{\begin{eqnarray}}
\newcommand\eeqn{\end{eqnarray}}
\newcommand\nn{\nonumber}
\newcommand\fc{\frac}
\newcommand\lt{\left}
\newcommand\rt{\right}
\newcommand\pt{\partial}
\newcommand\tx{\text}
\newcommand\mc{\mathcal}

\allowdisplaybreaks

\title{Static and spherically symmetric black holes in gravity with a background Kalb-Ramond field}
\author{
Ke Yang\footnote{keyang@swu.edu.cn, corresponding author}, 
Yue-Zhe Chen,
Zheng-Qiao Duan,
Ju-Ying Zhao
}

\affiliation{
School of Physical Science and Technology, Southwest University, Chongqing 400715, China}

\begin{abstract}

The Lorentz symmetry of gravity is spontaneously broken when the nonminimally coupled Kalb-Ramond field acquires a nonzero vacuum expectation value. In this work, we present exact solutions for static and spherically symmetric black holes in the framework of this Lorentz-violating gravity theory. In order to explore the physical implications of Lorentz violation, we analyze the thermodynamic properties of the obtained solutions and evaluate the impact of Lorentz violation on some classical gravitational experiments within the Solar System. Furthermore, the Lorentz-violating parameter is constrained by using the measured results of these experiments.

\end{abstract}







           


\maketitle



\section{Introduction}

As a fundamental concept in modern physics, Lorentz symmetry states that the laws of physics are the same in all inertial reference frames. Although Lorentz symmetry is widely believed to be a fundamental symmetry of the nature and has been confirmed by numerous experiments and observations, it has been revealed that Lorentz symmetry may be broken at some energy scale in plenty of theories, such as string theory \cite{Kostelecky1989a}, loop quantum gravity \cite{Alfaro2002}, Horava-Lifshitz gravity \cite{Horava2009a}, noncommutative field theory \cite{Carroll2001}, Einstein-\ae ther theory \cite{Jacobson2001}, massive gravity \cite{Dubovsky2005}, $f(T)$ gravity \cite{Bengochea2009}, very special relativity \cite{Cohen2006}, and so on. The study of Lorentz symmetry breaking (LSB) is important for understanding fundamental physical processes in high-energy physics and gravity. 

LSB can occur either explicitly or spontaneously. The explicit LSB occurs when the Lagrange density is not Lorentz invariant, i.e., the physical laws have different forms in certain reference frames. However, the spontaneous LSB occurs when Lagrange density is Lorentz invariant, but the ground state of a physical system does not exhibit Lorentz symmetry. A general framework for studying the spontaneous LSB is the Standard-Model extension \cite{Kostelecky2004a}. The simplest field theories proposed within this framework are known as bumblebee models \cite{Kostelecky1989a,Kostelecky1989,Kostelecky1989b,Bailey2006,Bluhm2008a}. In bumblebee models, a vector field known as the bumblebee field acquires a nonzero vacuum expectation value (VEV). The nonzero VEV  selects a specific direction, leading to the violation of particle local Lorentz invariance.

An exact solution for a static and spherically symmetric spacetime in the bumblebee gravity model was reported by Casana et al.~in Ref.~\cite{Casana2018}. This Schwarzschild-like solution has been extensively studied in various aspects, such as the gravitational lensing \cite{Ovgun2018}, Hawking radiation \cite{Kanzi2019}, accretion process \cite{Yang2019b,Cai2022a}, and quasinormal modes \cite{Oliveira2021}. Subsequently, Maluf et al.~obtained an (A)dS-Schwarzschild-like solution by relaxing the vacuum conditions \cite{Maluf2021}. Xu et al.~found new classes of static spherical bumblebee black holes by incorporating a background bumblebee field with a nonvanishing temporal component \cite{Xu2023}. The associated thermodynamic properties and observational implications were studied in Refs.~\cite{Mai2023,Xu2023a,Liang2023}. Rotating bumblebee black holes were investigated by Ding et al.~in Refs.~\cite{Ding2020a,Ding2021a}. The shadow \cite{Wang2022}, accretion process \cite{Liu2019}, quasinormal modes \cite{Liu2023}, and quasiperiodic oscillations \cite{Wang2022a} were discussed in these rotating bumblebee black holes. An exact rotating Ba\~nados-Teitelboim-Zanelli-like black hole solution was obtained in Ref.~\cite{Ding2023} and its quasinormal modes were studied analytically in Ref.~\cite{Chen2023}. A Schwarzschild-like black hole with a global monopole was presented in Ref.~\cite{Gullu2022}, and its quasinormal modes were analyzed in Refs.~\cite{Zhang2023,Lin2023}. Moreover, other black hole solutions in the framework of the bumblebee gravity model were investigated in Refs.~\cite{Ding2021,Jha2021,Ding2023a}. A traversable bumblebee wormhole solution was found in Ref.~\cite{Ovgun2019}. In addition, the gravitational waves in bumblebee gravity were analyzed in Refs.~\cite{Liang2022,Amarilo2023}.  

Instead of a vector field, a rank-two antisymmetric tensor field, called the Kalb-Ramond (KR) field, has also been considered as a source for LSB \cite{Altschul2010}. The KR field emerges in the spectrum of bosonic string theory \cite{Kalb1974}, and its properties have been extensively studied in various contexts, such as black hole physics \cite{Kao1996,Kar2003,Chakraborty2017}, cosmology \cite{Nair2022}, and the braneworld scenario \cite{Fu2012, Chakraborty2016}. When the KR field nonminimally couples to gravity and acquires a nonzero VEV, the Lorentz symmetry is spontaneously broken. An exact static and spherically symmetric solution was reported by Lessa et al.~under the VEV background of the KR field \cite{Lessa2020}. Subsequently, the motion of massive and massless particles in the vicinity of this static spherical KR black hole was studied in Ref.~\cite{Atamurotov2022}. The gravitational deflection of light and shadow cast by the rotating KR black holes were further investigated in Ref.~\cite{Kumar2020c}. Moreover, traversable wormhole solutions were constructed in this theory \cite{Lessa2021,Maluf2022}, and the implications of the VEV background on Bianchi type I cosmology were explored \cite{Maluf2022a}. 

In this work, we focus on the construction of new exact solutions for the static and spherically symmetric spacetime in the presence and absence of the cosmological constant under the nonzero VEV background of the KR field. The layout of the paper is as follows: In Sec.~\ref{Solution}, we solve the theory to obtain analytic solutions for the static and spherically symmetric spacetime, considering both the cases with and without the cosmological constant. In Sect.~\ref{Thermodynamics}, some basic thermodynamic properties of the obtained KR black holes are analyzed. In Sect.~\ref{Constraints}, we impose constraints on the theoretical parameter associated with Lorentz-breaking effects through classical Solar System tests. Finally, brief conclusions are given.

\section{KR Black Hole solutions}\label{Solution}

We start from the Einstein-Hilbert action nonminimally coupled to a self-interacting KR field in the form \cite{Altschul2010,Lessa2020}
\beqn
S \!&\!=\!&\! \fc{1}{2\kappa}\int{}d^4x\sqrt{-g}\Bigg[R-2\Lambda-\fc{1}{6}H^{\mu\nu\rho}H_{\mu\nu\rho}-V(B^{\mu\nu}B_{\mu\nu})   \nn\\
\!&\!+\!&\! \xi_2 B^{\rho\mu}B^{\nu}{}_\mu R_{\rho\nu}+\xi_3 B^{\mu\nu}B_{\mu\nu}R  \Bigg]+\int{d^4x\sqrt{-g}\mathcal{L}_\tx{M}},
\label{Main_Action}
\eeqn
where $\kappa=8 \pi G$ with $G$ the Newtonian constant of gravitation, $\xi_2$ and $\xi_3$ are the coupling constants between the gravity and the KR field,  $H_{\mu\nu\rho}\equiv \pt_{[\mu}B_{\nu\rho]}$ is the strength of the KR field, and $\Lambda$ is the cosmological constant. The self-interaction potential $V(B^{\mu\nu}B_{\mu\nu})$ depends on $B^{\mu\nu}B_{\mu\nu}$ in order to maintain the theory invariant upon observer local Lorentz transformations. As the cosmological constant $\Lambda$ is accounted for separately in the action, the potential is chosen to be zero at its minimum. 

By varying the action \eqref{Main_Action} with respect to the metric $g^{\mu\nu}$, one obtains the gravitational field equations
\beqn
R_{\mu \nu }-\frac{1}{2}g_{\mu \nu }R+\Lambda  g_{\mu \nu }= T^{\tx{KR}}_{\mu\nu}+T^{\tx{M}}_{\mu\nu},
\label{Modified_Einstein_EQ}
\eeqn
where $T^{\tx{M}}_{\mu\nu}$ is the energy-momentum tensor of matter fields, and
\beqn
T^{\tx{KR}}_{\mu\nu} \!&\!=\!&\!\frac{1}{2} H_{\mu \alpha \beta } H_{\nu }{}^{\alpha \beta } \!-\! \frac{1}{12} g_{\mu \nu } H^{\alpha \beta \rho } H_{\alpha \beta \rho }\!+\!2V'(X) B_{\alpha\mu}B^{\alpha}{}_\nu  \nn\\
\!&\!-\!&\! g_{\mu\nu}V(X)+ \xi_2 \bigg[\frac{1}{2} g_{\mu \nu } B^{\alpha \gamma } B^{\beta }{}_{\gamma }R_{\alpha \beta } - B^{\alpha }{}_{\mu } B^{\beta }{}_{\nu }R_{\alpha \beta }\nn\\
\!&\!-\!&\!  B^{\alpha \beta } B_{\nu \beta } R_{\mu \alpha }-B^{\alpha \beta } B_{\mu \beta } R_{\nu \alpha }+\frac{1}{2} \nabla _{\alpha }\nabla _{\mu }\left(B^{\alpha \beta } B_{\nu \beta }\right)
\nn\\
\!&\!+\!&\!
\frac{1}{2} \nabla _{\alpha }\nabla _{\nu }\left(B^{\alpha \beta } B_{\mu \beta }\right)-\frac{1}{2}\nabla ^{\alpha }\nabla _{\alpha }\left(B_{\mu }{}^{\gamma }B_{\nu \gamma } \right)
\nn\\
\!&\!-\!&\! \frac{1}{2} g_{\mu \nu } \nabla _{\alpha }\nabla _{\beta }\left(B^{\alpha \gamma } B^{\beta }{}_{\gamma }\right)\bigg],
\label{EoM1}
\eeqn
where the prime represents the derivative with respect to the argument of the corresponding functions. With the Bianchi identities, it is clear that the total energy-momentum tensor $T_{\mu\nu}^\tx{tot}=T^{\tx{KR}}_{\mu\nu}+T^{\tx{M}}_{\mu\nu}$ is conserved.

In order to generate a nonvanishing VEV for the KR field, $\langle B_{\mu\nu} \rangle=b_{\mu\nu}$, we assume a potential with a general form given by $V=V\lt(B^{\mu\nu} B_{\mu\nu}\pm b^2\rt)$, where the sign $\pm$ is chosen such that $b^2$ is a positive constant \cite{Bluhm2008,Altschul2010,Lessa2020}. Consequently, the VEV is determined by the constant norm condition $b^{\mu\nu} b_{\mu\nu}=\mp b^2$.  Upon the vacuum condensation, the gauge invariance $B_{\mu\nu} \to B_{\mu\nu}+\pt_{[\mu}\Gamma_{\nu]}$ of the KR field is spontaneously broken. Due to the nonminimal coupling of the KR field to gravity, the symmetry-breaking VEV background leads to the violation of particle local Lorentz invariance. Furthermore, the term $\xi_3 B^{\mu\nu}B_{\mu\nu}R$ in the action \eqref{Main_Action} transforms to $\mp\xi_3 b^2 R$ in the vacuum, which can be absorbed into Einstein-Hilbert terms through a redefinition of variables. 

It is convenient to decompose the antisymmetric tensor $B_{\mu\nu}$ to be $B_{\mu\nu}=\tilde E_{[\mu}v_{\nu]}+\epsilon_{\mu\nu\alpha\beta}v^\alpha \tilde B^\beta$ with $v^\alpha$ a timelike 4-vector \cite{Altschul2010,Lessa2020}. The pseudofields $\tilde E_\mu$ and $\tilde B_\mu$ are spacelike, satisfying $\tilde E_\mu v^\mu=\tilde B_\mu v^\mu=0$. Analogous to Maxwell electrodynamics, these fields can be interpreted as the pseudoelectric and pseudomagnetic fields, respectively. Assuming that the only nonvanishing terms are $b_{10}=-b_{01}=\tilde E(r)$ in the VEV, or equivalently, $\bold{b}_2=-\tilde E(r)dt\wedge dr$ in terms of differential forms, it follows that the vacuum field exhibits a pseudoelectric configuration. Consequently, this configuration automatically vanishes the KR field strength, i.e., $H_{\lambda\mu\nu} = 0$ or $\bold{H}_3=d \bold{b}_2=0$.

Here, our focus is on investigating a static and spherically symmetric spacetime under the nonzero VEV background of the KR field. The metric ansatz is given by
\beq
ds^2=-A(r)dt^2+B(r)dr^2+r^2 d\theta^2+r^2 \sin^2\theta d\phi^2.
\eeq
Correspondingly, the pseudoelectric field $\tilde E(r)$ can be rewriten as $\tilde E(r)=|b|\sqrt{A(r)B(r)/2}$, such that the constant norm condition $b^{\mu\nu}b_{\mu\nu}=-b^2$ is satisfied. 

Under the VEV configuration, it is advantageous to reformulate the field equation as
\beqn
R_{\mu\nu} \!&\!=\!&\! \Lambda  g_{\mu \nu }+V' \left(2 b_{\mu \alpha } b_{\nu }{}^{\alpha }+b^2 g_{\mu \nu }\right)  +\xi_2 \bigg[g_{\mu \nu } b^{\alpha \gamma }  b^{\beta }{}_{\gamma }R_{\alpha \beta }\nn\\
\!&\!-\!&\! b^{\alpha }{}_{\mu } b^{\beta }{}_{\nu }R_{\alpha \beta }-b^{\alpha \beta } b_{\mu \beta } R_{\nu \alpha }- b^{\alpha \beta } b_{\nu \beta } R_{\mu \alpha } \nn\\
\!&\! + \!&\!
\frac{1}{2} \nabla _{\alpha }\nabla _{\mu }\left(b^{\alpha \beta } b_{\nu \beta }\right)+\frac{1}{2} \nabla _{\alpha }\nabla _{\nu }\lt(b^{\alpha \beta } b_{\mu \beta }\rt)\nn\\
\!&\!-\!&\! \frac{1}{2}\nabla ^{\alpha }\nabla _{\alpha }\lt(b_{\mu }{}^{\gamma }b_{\nu \gamma } \rt)\bigg].
\label{EoM2}
\eeqn
With the metric ansatz, the field equations \eqref{EoM2} can be written explicitly as
\begin{subequations}
\beqn
\frac{2 A''}{A}-\fc{A'}{A}\frac{ B'}{ B}-\frac{A'^2}{A^2}+\fc{4}{r}\frac{A'}{A}
+\frac{4 \Lambda  }{1-\ell}B \!&\!=\!&\! 0,\label{EoM_1}\\
\frac{2 A''}{A}-\fc{A'}{A}\frac{ B'}{ B}-\frac{A'^2}{A^2}-\fc{4}{r}\frac{B'}{B}+\frac{4 \Lambda  }{1-\ell}B \!&\!=\!&\!0,\label{EoM_2}\\
\frac{2 A''}{A}-\frac{A' B'}{A B}-\frac{A'^2}{A^2}+\frac{1+\ell}{\ell  r}\left(\frac{A'}{A}-\frac{B'}{B}\right)\nn\\
-\left(1-\Lambda r^2-b^2 r^2 V'\right)\frac{2 B }{\ell r^2}+\frac{2 (1-\ell )}{\ell  r^2}  \!&\!=\!&\!0,\label{EoM_3}
\eeqn
\label{EOM_Exp}
\end{subequations}
where $\ell\equiv \xi_2b^2/2$.

\subsection{Absence of cosmological constant $\Lambda=0$}

First, by considering the spacetime without the cosmological constant, we would like to construct a Schwarzschild-like black hole in the theory. In this case, we take the assumption that $V'=0$, which corresponds to the case where the VEV is located at the local minimum of the potential. For instance, it can be simply realized by a potential of quadratic form, $V=\fc{1}{2}\lambda X^2$, with $X \equiv B^{\mu\nu} B_{\mu\nu}+ b^2$ and $\lambda$ a coupling constant \cite{Bluhm2008}.

By subtracting Eq.~\eqref{EoM_2} from Eq.~\eqref{EoM_1}, the following relation is obtained
\beqn
\frac{A'}{A}+\frac{B'}{B}=0.
\eeqn
It simply yields
\beq
A(r)=\fc{1}{B(r)}.
\label{Relation_AB}
\eeq
After subtracting Eq.~\eqref{EoM_3} from Eq.~\eqref{EoM_1} and substituting the relation \eqref{Relation_AB} into it, one obtains
\beq
A(r)=\frac{1}{1-\ell}-\frac{c_1}{r},
\eeq
where the integration constant $c_1$ can be determined using the Komar integral. 

Utilizing the time-translation Killing vector $K^\mu = (1,0,0,0)$, the current can be constructed as $J^\mu=K_\nu R^{\mu\nu}$. From the modified Einstein equations \eqref{Modified_Einstein_EQ}, the current can be expressed as $J^\mu=K_\nu \lt(T^{\mu\nu}_\tx{tot}-\fc{1}{2}g^{\mu\nu}T_\tx{tot}+\Lambda g^{\mu\nu}\rt)$ as well.  It can be observed that the only distinction from general relativity is the replacement of the energy-momentum tensor of the matter fields $T^\tx{M}_{\mu\nu}$ with the total energy-momentum tensor $T_{\mu\nu}^\tx{tot}$. With the Bianchi identity and the Killing's equation, it can be proven that the current is conserved, i.e., $\nabla_\mu J^\mu=\nabla_\mu\lt(K_\nu R^{\mu\nu}\rt)=\lt(\nabla_\mu K_\nu\rt) R^{\mu\nu}+ \fc{1}{2}K_\nu\lt(\nabla^\nu R\rt)=0 $ \cite{Carroll2019}. By using the relation $\nabla_\mu \nabla_\nu K^\mu=K^\mu R_{\mu\nu}$ of the Killing vector, the current can be further written as a total derivative, $J^\mu=K_\nu R^{\mu\nu}=\nabla_\nu \nabla^\mu K^\nu$. By applying Stokes's theorem, the Komar mass $M$ can be calculated as
\beqn
M  \!&\!=\!&\! \frac{1}{4 \pi} \int_{\Sigma}\! d^3x \sqrt{\gamma^{(3)}} n_\mu J^\mu \!=\! \frac{1}{4 \pi}  \int_{\Sigma}\! d^3x \sqrt{\gamma^{(3)}} n_\mu \nabla_\nu \nabla^\mu K^\nu \nn\\
   \!&\!=\!&\!\frac{1}{4 \pi} \int_{\pt \Sigma} d^2x\sqrt{\gamma^{(2)}} n_\mu \sigma_\nu \nabla^\mu K^\nu 
 \nn\\
  \!&\!=\!&\! - \frac{1}{4 \pi} \int_{\pt \Sigma} d^2x\sqrt{\gamma^{(2)}}\nabla^{t}K^r= \fc{c_1}{2},
\eeqn
where $\Sigma$ represents a three-dimensional spacelike hypersurface with the unit normal vector $n_\mu=\lt(-\sqrt{A(r)},0,0,0\rt)$, and $\pt\Sigma$ denotes the boundary of $\Sigma$, which is a two-sphere at infinity with the unit normal vector $\sigma_\mu=\left(0,1/\sqrt{A(r)},0,0\right)$ and the induced metric $\gamma^{(2)}_{ij}=r^2 \lt (d\theta^2+\sin^2\theta d\phi^2 \rt)$.

Thus, the metric function $A(r)$ is determined as
\beq
A(r)=\frac{1}{1-\ell}-\frac{2M}{r}.
\label{Sol_A} 
\eeq 
It is straightforward to verify that this set of solutions \eqref{Relation_AB} and \eqref{Sol_A} satisfies all the equations of motion \eqref{EOM_Exp}. Consequently, a Schwarzschild-like metric is obtained \footnote{Note that the solution we obtained differs from that presented in the Ref.~\cite{Lessa2020}.}
\beqn
ds^2 \!&\!=\!&\! -\lt(\frac{1}{1-\ell}-\frac{2M}{r}\rt)dt^2+\fc{dr^2}{\frac{1}{1-\ell}-\frac{2M}{r}}+r^2 d\theta^2\nn\\
\!&\!+\!&\! r^2 \sin^2\theta d\phi^2.
\label{BH_without_CC}
\eeqn
The dimensionless parameter $\ell$ characterizes the effect of Lorentz violation caused by the nonzero VEV of the KR filed on spacetime. Due to strong constraints from experimental observations on Lorentz-violating effects in gravitational fields, which will be discussed later, the Lorentz-violating parameter $\ell$ is supposed to be very small. Note that, due to the nonvanishing of the Riemann tensor in the limit $r\to\infty$, the spacetime is not asymptotically Minkowski in this case. Furthermore, the Kretschmann scalar for the current spacetime is given by	
\beq
R^{\alpha\beta\gamma\delta}R_{\alpha\beta\gamma\delta}=\frac{48 M^2}{r^6}-\frac{16 \ell M}{(1-\ell) r^5}+\frac{4 \ell^2}{(\ell-1)^2 r^4}.
\eeq
This implies that the Lorentz-violating effects cannot be eliminated by mere coordinate transformations.

It is interesting that the black hole described by the metric \eqref{BH_without_CC} possesses a horizon radius $r_\tx{h}=2(1-\ell) M$, which is shifted by the Lorentz-violating parameter $\ell$. This is different from the case of a bumblebee black hole, where the horizon radius is identical to that of a Schwarzschild black hole \cite{Casana2018}. At the horizon, the Kretschmann scalar is finite, given by $R^{\alpha\beta\gamma\delta}R_{\alpha\beta\gamma\delta}=\frac{3-(2-\ell) \ell}{4 (1-\ell)^6 M^4}$, indicating that the singularity at the horizon can be removed through a coordinate transformation. However, the Kretschmann scalar diverges at the center $r=0$, signifying an intrinsic and nonremovable singularity at that point.

\subsection{Presence of cosmological constant $\Lambda \neq 0$}

When the cosmological constant is present, it is found that the assumption $V'=0$ does not support a self-consistent solution that satisfies all the equations of motion. Therefore, following the approach in Ref.~\cite{Maluf2021}, the vacuum condition is relaxed to be $V=0$ but $V'\neq 0$. The most commonly considered potential form that satisfies the condition is a linear form, given by $V=\lambda X$, where $\lambda$ is a Lagrange multiplier field \cite{Bluhm2008}. Consequently, the derivative of the potential with respect to $X$ is $V'(X)=\lambda$. The equation of motion of the Lagrange-multiplier $\lambda$ restricts the theory to the extrema of the potential, satisfying $X=0$, such that $b_{\mu\nu}$ is the VEV of the KR field for the on-shell $\lambda$. Note that the off-shell value of the  Lagrange multiplier should have the same sign as $X$ to keep the positivity of the potential $V$ \cite{Bluhm2008}. In principle, the Lagrange-multiplier field $\lambda$ can also be expanded around its vacuum value, i.e., $\lambda=\langle \lambda \rangle + \tilde\lambda$, and $\langle \lambda \rangle$ could vary with spacetime position. However, it is convenient to fix $\tilde \lambda=0$ by choosing the initial conditions and to assume that $\langle \lambda \rangle$ is a real constant, then the on-shell value $\lambda \equiv \langle \lambda \rangle$ is determined by the field equations completely \cite{Bluhm2008,Maluf2021}.

Now, by subtracting Eq.~\eqref{EoM_2} from Eq.~\eqref{EoM_1}, one obtains the same relation as in the previous case,
\beq
A(r)=\fc{1}{B(r)}.
\label{Relation_AB_2}
\eeq
Furthermore, by subtracting Eq.~\eqref{EoM_3} from Eq.~\eqref{EoM_1} and substituting the relation \eqref{Relation_AB_2} into it, one arrives at
\beq
A(r)=\frac{1}{1-\ell}-\frac{2M}{r}-\frac{(1-3 \ell)\Lambda + (1-\ell) b^2 \lambda}{3 (1-\ell)^2}r^2.
\label{Sol_A_2}
\eeq
Finally, by substituting Eqs.~\eqref{Relation_AB_2} and \eqref{Sol_A_2} into \eqref{EoM_1}, the on-shell value of $\lambda$ is determined by
\beq
\lambda = \frac{2 \ell \Lambda}{(1-\ell) b^2}.
\eeq
It is evident that the theory supports a solution with a nonvanishing cosmological constant if only $V'(X)=\lambda\neq 0$. 

After inserting the on-shell value of $\lambda$ into \eqref{Sol_A_2}, an (A)dS-Schwarzschild-like metric is obtained
\beqn
ds^2 \!&\!=\!&\! -\lt(\frac{1}{1-\ell}-\frac{2M}{r}-\frac{\Lambda r^2}{3\lt(1-\ell\rt)} \rt)dt^2\nn\\
\!&\!+\!&\! \fc{dr^2}{\frac{1}{1-\ell}\!-\!\frac{2M}{r}\!-\!\frac{\Lambda r^2}{3\lt(1-\ell\rt)} }+r^2 d\theta^2+r^2 \sin^2\theta d\phi^2.
\label{BH_with_CC}
\eeqn
The solution reduces to the Schwarzschild-like metric \eqref{BH_without_CC} when the cosmological constant vanishes. Likewise, it degenerates to the (A)dS-Schwarzschild metric when the Lorentz-violating parameter $\ell$ is set to zero. Furthermore, it is easy to demonstrate that the relation $R_{\mu\nu\rho\sigma}=\fc{\Lambda_\tx{eff}}{3}(g_{\mu\rho}g_{\nu\sigma}-g_{\mu\sigma}g_{\nu\rho})$ holds in the limit $r\to\infty$, where the effective cosmological constant reads $\Lambda_\tx{eff}\equiv\fc{\Lambda}{1-\ell}$. It indicates that the spacetime approaches (A)dS at infinity. This behavior is also evident from the metric \eqref{BH_with_CC}, where the metric functions approximate $A(r)=1/B(r)\to-{r^2 \Lambda_\tx{eff} }/{3}$ as $r\to \infty$ tends to infinity, exhibiting the same asymptotic behavior as the (A)dS metric. Note that the truly contributing cosmological constant is the effective one $\Lambda_\tx{eff}\equiv\Lambda / (1-\ell)$ due to the additional contributions from the modified Einstein equations when $V'(X)$ is nonzero.

\begin{figure}[t]
\begin{center}
\includegraphics[width=7cm]{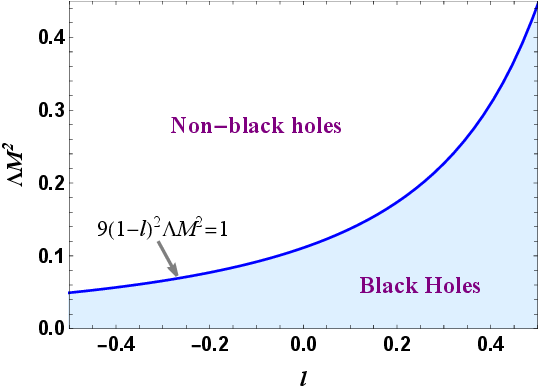}
\end{center}
\caption{The parameter space $(l,\Lambda M^2)$ for black hole solutions and nonblack hole solutions, where the light blue region represents the parameter space of black hole solutions.}
\label{Parameter_Space}
\end{figure}

When the cosmological constant is negative, $\Lambda<0$, the event horizon exists for any parameter, and thus the metric \eqref{BH_with_CC} always supports black hole solutions in this case. However, as shown in Fig.~\ref{Parameter_Space}, when the cosmological constant is positive, $\Lambda>0$, the black hole solutions exist only for the parameters $\ell$ and $\Lambda M^2$ satisfying $9(1-\ell)^2\Lambda M^2\leq 1$, where the equality sign represents the case when the event horizon and the cosmological horizon coincide with each other.

\section{Thermodynamics}\label{Thermodynamics}

Research on black hole thermodynamics has shown that they not only possess the four laws of black hole thermodynamics \cite{Bardeen1973,Kastor2009}, but they also exhibit phase structures akin to that of ``every day thermodynamics" \cite{Kubiznak2017,Wei2015a,Yang2022c}. This fascinating discovery provides insights into the nature of black holes and gravity. Asymptotically AdS black holes are wellknown for their close relationship with AdS/CFT duality and diverse thermodynamic behaviors, including various phase transitions. However, the presence of two distinct temperatures in asymptotically dS black holes renders the system in a nonequilibrium state. Although some attempts have been made to investigate the thermodynamics of these black holes \cite{Mbarek2019}, our understanding in this area remains limited. For the sake of simplicity, we primarily focus on analyzing some fundamental thermodynamic properties of the obtained AdS-Schwarzschild-like black hole in this section.

By solving $A(r_\tx{h})=0$ in Eq.~\eqref{BH_with_CC}, the black hole mass can be expressed with the radius of the event horizon $r_\tx{h}$, 
\beq
M=\frac{\left(3-\Lambda r_\tx{h}^2\right)r_\tx{h}}{6 (1-\ell)}.
\label{Relation_M_rh}
\eeq
Since we consider a static spherical spacetime with an effective cosmological constant $\Lambda_\tx{eff}$, the  black hole mass $M$ is associated with the black hole's enthalpy. In this case, the effective cosmological constant plays the role of a thermodynamic pressure, given by $P=-\fc{\Lambda_\tx{eff}}{8\pi}=-\fc{\Lambda}{8\pi(1-l)}$.  Considering that the Lorentz-violating parameter $\ell$ is a dimensionless constant, the enthalpy can be expressed as a function of entropy $S$ and pressure $P$, i.e., $M=M\lt(S,P\rt)$. Consequently, the first law of the AdS-Schwarzschild-like black hole reads
\beq
dM=\mathcal{T}dS+\mathcal{V}dP,
\label{First_law}
\eeq
where $\mathcal{T}$ is the Hawking temperature and $\mathcal{V}$ the thermodynamic volume.

By employing the metric \eqref{BH_with_CC} and Eq.~\eqref{Relation_M_rh}, the Hawking temperature is given by
\beq
\mathcal{T}=-\lt.\fc{1}{4\pi}\fc{\pt g_{tt}}{\pt r}\rt|_{r_\tx{h}}=\frac{1-\Lambda  r_\tx{h}^2}{\lt(1-\ell \rt)4\pi r_\tx{h} }.
\label{Temperature}
\eeq
It is observed that the Lorentz-violating parameter $\ell$ has an impact on the temperature of the black hole. Specifically, the temperature is higher compared to that of an AdS-Schwarzschild black hole for a positive $\ell$, while it is lower for a negative $\ell$. 

Further, by utilizing the first law \eqref{First_law} and the Hawking temperature \eqref{Temperature}, the entropy can be obtained as
\beqn
\!\!S=\!\int{}\lt(\fc{d M}{\mathcal{T}} \rt)_P\!=\!\int{}\fc{1}{\mathcal{T}}\lt(\fc{\pt M}{\pt r_\tx{h}} \rt)_P d r_\tx{h}=\pi r_\tx{h}^2=\fc{A_\tx{h}}{4},
\eeqn
where $A_\tx{h}=4\pi r_\tx{h}^2$ is the horizon area. It is evident that the entropy of the black hole still satisfies the Bekenstein-Hawking area-entropy relation.

Correspondingly, the thermodynamic volume is calculated as 
\beq
\mathcal{V}=\lt(\fc{\pt M}{\pt P}\rt)_S=\lt(\fc{\pt M}{\pt \Lambda }\rt)_S\lt(\fc{\pt \Lambda}{\pt P }\rt)_S=\frac{4 \pi r_\tx{h}^3}{3}.
\eeq

Further, utilizing the aforementioned results, the Smarr formula is obtained as
\beq
M=2\lt(\mathcal{T}S-\mathcal{V}P \rt).
\eeq 
Therefore, the fist law and Smarr formula are identical to the results obtained for the AdS-Schwarzschild black hole.

The specific heat is useful for analyzing the local stability of a black hole solution. Specifically, a black hole is stable when its specific heat is positive, while it is unstable when its specific heat is negative. The specific heat is computed by
\beqn
\!C_P\!=\!\!\lt(\fc{\pt M}{\pt \mathcal{T}}\rt)_P\!\!\!=\!\lt(\fc{\pt M}{\pt r_\tx{h} }\rt)_P\!\!\lt(\fc{\pt r_\tx{h}}{\pt \mathcal{T} }\rt)_P\!\!\!=\! 2 \pi  r_\tx{h}^2 \!\lt(\frac{ \Lambda  r_\tx{h}^2\!-\!1}{\Lambda  r_\tx{h}^2 \!+\! 1}\rt)\!.
\eeqn
Interestingly, although both the mass and temperature of the black hole contain the Lorentz-violating parameter $\ell$, the specific heat is independent of it. Consequently, the AdS-Schwarzschild-like black hole is locally stable only if its radius is sufficiently large, satisfying $r_\tx{h}>\sqrt{-1/\Lambda}$. However, for the Schwarzschild-like black hole with $\Lambda=0$, the negative specific heat indicates its locally instability.

Furthermore, a black hole is globally stable if its Gibbs free energy is negative, but it is globally unstable if its Gibbs free energy is positive. Hence, in order to investigate the global stability of the black hole, we further evaluate its Gibbs free energy in the canonical ensemble, which is expressed as
\beq
F=M-\mathcal{T}S=\frac{\lt(3+ \Lambda r_\tx{h}^2\rt)r_\tx{h}}{12(1-\ell)}.
\eeq
It is evident that the free energy is negative only when $r_\tx{h}>\sqrt{-3/\Lambda}$, indicating that only large AdS-Schwarzschild-like black holes are globally stable. For the Schwarzschild-like black hole, its free energy is always positive and hence it is globally unstable. It is worth noting that despite the presence of the Lorentz-violating parameter $\ell$ in the free energy, it does not affect the critical size that determines the stability of the black holes.

\section{Solar system tests}\label{Constraints}

General relativity (GR) has undergone extensive testing, especially in the weak field regime, such as within the Solar System. So far, no evidence contradicting the theory has been found. However, it would be helpful to evaluate the impact of the present Lorentz-violating theory by testing the derived solutions within the Solar System. The theoretical parameter $\ell$ can be constrained by comparing the experimental results with the predictions of the theory. Since the influence of the cosmological constant can be neglected at the scale of the Solar System, we focus solely on considering the Schwarzschild-like metric \eqref{BH_without_CC}.

The motion of a test particle along its geodesics can be described by the Lagrangian
\beq
\mathcal{L}=-\fc{1}{2}g_{\alpha\beta}\dot{x}^\alpha\dot{x}^\beta,
\label{Lagrangian_Abst}
\eeq
where the dot represents the derivative with respect to the affine parameter $\lambda$. From the normalization conditions for the four-velocity of timelike and null particles, one can express the Lagrangian density as $\mathcal{L}=\eta/2$. Accordingly, for massless photons, $\eta=0$, while for massive particles, $\eta=1$ when the affine parameter $\lambda$ is chosen to be the proper time $\tau$.

By substituting the metric \eqref{BH_without_CC} into the Lagrangian \eqref{Lagrangian_Abst}, one obtains
\beqn
\mathcal{L}=\frac{1}{2}\bigg(A(r) \dot{t}^{2} - A(r)^{-1} \dot{r}^{2} -r^{2} \dot{\theta}^{2}-r^{2}\sin^{2}\theta\dot{\phi}^{2}\bigg).
\eeqn
The Euler-Lagrange equation corresponding to $\theta$ is given by
\beq
r^2 \ddot\theta+2r\dot{r}\dot\theta-r^2\dot{\phi}^2\sin\theta\cos\theta=0.
\eeq
With the initial condition $\theta=\fc{\pi}{2}$ and $\dot{\theta}=0$, one has $\ddot\theta=0$. It indicates that the particle will be restricted to motion within the equatorial plane. In this case, the Lagrangian density \eqref{Lagrangian_Abst} is rewritten as 
\beq
A(r) \lt(\fc{dt}{d\lambda}\rt)^{2} - A(r)^{-1} \lt(\fc{dr}{d\lambda}\rt)^{2}-r^{2} \lt(\fc{d\phi}{d\lambda}\rt)^{2}=\eta.
\label{Lagrangian_Expl}
\eeq

Since the spacetime is static and spherically symmetric, the Lagrangian is independent of $t$ and $\phi$. As a result, there are two conserved quantities: the energy $E$ and the angular momentum $L$, given by
\beqn
E &=& \fc{\pt \mathcal{L}}{\pt \dot{t} }=A(r)\fc{dt}{d\lambda}, \label{Energy_Cons}\\
L &=& -\fc{\pt \mathcal{L}}{\pt \dot{\phi}}=r^{2}\fc{d\phi }{d \lambda}.\label{Momentum_Cons}
\eeqn

Now that we have all the required equations, we can proceed to examine three classical tests within the Solar System: the perihelion precession of Mercury, the deflection of light by the Sun, and the Shapiro time delay.

\subsection{Perihelion precession of Mercury}

Mercury can be viewed as a test particle, allowing us to choose the affine parameter $\lambda$ as the proper time $\tau$. By utilizing Eqs.~\eqref{Lagrangian_Expl}, \eqref{Energy_Cons} and \eqref{Momentum_Cons}, the orbital equation is derived as
\beqn
\lt[\frac{d}{d\phi}\lt(\frac{1}{r}\rt)\rt]^2=\frac{E^2}{L^2}-\frac{A(r) }{L^2}\left(1+\frac{L^2}{r^2}\right).
\eeqn 
After redefining $u=\fc{L^2}{Mr}$ and differentiating the equation with respect to $\phi$, we have 
\beq
\fc{d^2 u}{d\phi^2}+\fc{u}{1-\ell}-1=\frac{3M^2}{L^2} u^2.
\label{Orbital_equation}
\eeq
It is observed that the coefficient of the second term has been modified by the Lorentz-violating parameter $\ell$, which is supposed to be very small, $\ell\ll 1$. The equation can be solved using a perturbative method. Noting that the last term, represented as $\frac{3G^2M^2}{c^4L^2}$ in the International System of Units, is a first-order small quantity and would be absent in the Newtonian limit \cite{Carroll2019}. Therefore, by expanding the exact solution $u$ as $u \approx u_{0}+u_{1}$, where $u_{0}$ and $u_{1}$ are respectively the zeroth-order and first-order approximations, satisfying \cite{Carroll2019}
\beqn
\fc{d^{2}u_{0}}{d\phi^{2}}+\frac{u_{0}}{1-\ell}-1&=&0,\label{Zero_Order}\\
\fc{d^{2}u_{1}}{d\phi^{2}}+\frac{u_{1}}{1-\ell}&=&\frac{3M^2}{L^2}u_{0}^{2}. \label{First_Order}
\eeqn

The solution of zeroth-order equation \eqref{Zero_Order} is given by
\beq
u_{0}=(1-\ell)\left[1+e\cos\left(\frac{\phi}{\sqrt{1-\ell}}\right)\right],
\eeq
where the integration constants are chosen such that the initial value is $\phi_0=0$ and the orbital eccentricity is $e$. If $\ell=0$, this solution describes a closed ellipse, which is precisely the orbit predicted by Newtonian mechanics. However, due to the appearance of the Lorentz-violating parameter $\ell$ in the phase, the elliptical orbit is not closed any more.

By substituting the zeroth-order solution into Eq.~\eqref{First_Order}, it is straightforward to check that the first-order solution is given by
\beqn
u_{1}\!&\!=\!&\!\frac{3 (1-\ell)^3 M^2}{L^2}\lt[\lt(1+\frac{e^2}{2}\rt)-\frac{e^2}{6} \cos \left(\frac{2 \phi }{\sqrt{1-\ell}}\right)\rt.\nn\\
\!&\!+\!&\! \lt.\frac{e \phi}{\sqrt{1-\ell}}\sin \left(\frac{\phi }{\sqrt{1-\ell}}\right)\rt].
\eeqn
It is noted that the first term in the square bracket is simply a constant displacement, and the second term oscillates around zero. However, the  third term accumulates over successive orbits and contributes the important effect during rotation. We therefore only consider the contribution of this term in the first-order solution. 

Combining the zeroth-order and first-order solutions, we finally arrive at
\beqn
u\!&\!\approx \!&\! (1-\ell)\left[1+e\cos\lt(\frac{\phi}{\sqrt{1-\ell}}\right)\rt.\nn\\
\!&\!+\!&\! \lt.\fc{3(1-\ell)^\fc{3}{2}M^2}{L^2}e\phi \sin \left(\frac{\phi }{\sqrt{1-\ell}}\right) \rt].
\eeqn
For convenience, it can be further approximated and rewritten in the form of an elliptical equation,
\beq
u\!\approx \! (1\!-\!\ell)\left[1\!+\!e\cos\left[\!\lt(1\!-\!\frac{3 (1-\ell)^2 M^2}{L^2}\rt)\!\frac{\phi}{\sqrt{1-\ell}}\right]\right].
\eeq
The period of $\phi$ of the orbits reads
\beq
\Phi\!=\!\frac{2 \pi  \sqrt{1-\ell}}{1\!-\!\frac{3 (1-\ell)^2 M^2}{L^2}}\!\approx\! 2 \pi  \sqrt{1-\ell}\lt(1+\fc{3 (1\!-\!\ell)^2 M^2}{L^2} \rt).
\eeq
During each orbit of the planet, the perihelion advances by an angle $\Delta\Phi=\Phi-2\pi$, which is obtained by taking the lowest order in the expansion of $\ell$, yielding
\beq
\Delta\Phi\approx \fc{6\pi M^2}{L^2}-\ell \pi.
\eeq
The first term $\Delta_\tx{GR}=6\pi\fc{M^2}{L^2}$ is exactly the estimation derived from GR, while the second term $\Delta_\tx{LV}=-\ell \pi$ represents the contribution arising from the effect of LSB.
 
For Mercury, the predicted value of its perihelion precession in GR is $ 42.9814^{\prime \prime}/\tx{century}$, while the observed value is measured as $\Delta\Phi_\tx{Ex}=(42.9794 \pm 0.0030)^{\prime \prime}/\tx{century}$ \cite{Casana2018}. The precession value can be converted into the precession per orbit by using the period of Mercury, which is approximately 87.969 days. By requiring that the theoretical prediction of the present theory is consistent with the experimental value, i.e., $\Delta\Phi_\tx{Ex}^\tx{Min}\leq \Delta\Phi \leq \Delta\Phi_\tx{Ex}^\tx{Max}$, we obtain the constraints on the Lorentz-violating parameter as $-3.7 \times 10^{-12}\leq \ell \leq 1.9\times10^{-11}$.

\subsection{Deflection of light}

For photons with $\eta=0$, the orbital equation can be obtained from Eqs.~\eqref{Lagrangian_Expl}, \eqref{Energy_Cons}, and \eqref{Momentum_Cons}, given by
\beqn
\lt[\frac{d}{d\phi}\lt(\frac{1}{r}\rt)\rt]^2=\frac{E^2}{L^2}-\frac{A(r) }{r^2}.
\eeqn
By redefining $u={1}/{r}$ and differentiating the equation with respect to $\phi$, we have 
\beq
\fc{d^2 u}{d\phi^2}+\fc{u}{1-\ell}=3M u^2.
\label{Orbital_equation}
\eeq
Following the approach in the previous subsection, we employ a perturbative method to solve the equation and expand the exact solution $u$ as $u \approx u_{0}+u_{1}$. Once again, the last term, represented as $\frac{3GM}{c^2}u^2$ in the International System of Units, is a small first-order quantity and therefore can be neglected in the zeroth-order calculation. As a result, the zeroth-order solution satisfies a homogeneous equation, i.e.,
\beq
\frac{d^{2}{{u}_{0}}}{d\phi^{2}} + \frac{{u}_{0}}{1-\ell}=0.
\eeq
The solution is given by
\beq
{u}_{0}=b^{-1}\sin\left[\frac{\phi}{\sqrt{1-\ell}}\right],
\label{Light_0th}
\eeq
where the integration constants have been chosen appropriately so that $b$ represents the impact parameter and the incident angle is $\phi_0=0$. When $\ell=0$, the solution describes a straight line, which is exactly the Newtonian prediction. However, in the presence of Lorentz-violating effect, it is evident that the exit angle of the light is no longer $\phi=\pi$, but is instead given by $\phi=\sqrt{1-\ell}\pi\approx \pi -\fc{\pi \ell}{2}$. Consequently, the Lorentz-breaking effect will induce an additional deflection angle of $\delta_\tx{LV}=-\fc{\pi \ell}{2}$.

By substituting the zeroth-order solution into the first-order equation 
\beq
u_1''+\frac{u_1}{1-\ell}=3 M u_0^2,
\eeq
we obtain the solution
\beq
{u}_{1}=\frac{(1-l) M}{b^2}\lt[1+\cos^{2}\lt(\fc{\phi}{\sqrt{1-\ell}}\rt)\rt].
\label{Light_1th}
\eeq
Therefore, we arrive at the final solution
\beq
u\!=\!\fc{1}{b}\sin\left[\frac{\phi}{\sqrt{1\!-\!\ell}}\right]\!+\!\frac{(1\!-\!\ell) M}{ b^2}\lt[1\!+\!\cos^{2}\lt(\fc{\phi}{\sqrt{1\!-\!\ell}}\rt)\rt].
\label{Solution_u}
\eeq

When a ray of light is incident from infinity and exits to infinity, we have $r\to\infty$ or $u\to0$. Then, by solving the above equation with $u=0$, we obtain the expression
\beq
\sin\left[\frac{\phi}{\sqrt{1-\ell}}\right]=\frac{b-\sqrt{b^2+8(1-\ell)^2 M^2}}{2(1-\ell) M}.
\eeq
Taking into account $\ell\ll 1$ and $M \ll 1$, the incident angle at lowest order can be approximated as 
\beq
\phi_\tx{in}=-\frac{2 M}{b}.
\eeq
Furthermore, the exit angle at the lowest order can be expressed as
\beq
\phi_\tx{ex}=\pi+\frac{2 M}{b}-\frac{\pi  \ell}{2}.
\eeq
Here, the second term precisely represents the Lorentz-violating effect observed in the zeroth-order solution, as discussed earlier. 

So the total angle of deflection is obtained as
\beq
\delta=\frac{4 M}{b}-\frac{\pi \ell}{2}.
\eeq
The first term, $\delta_\tx{GR}=\frac{4 M}{b}$, arises from the prediction of GR, while the second term, $\delta_\tx{LV}=-\frac{\pi \ell}{2}$, stems from the effect of LSB.

For a light ray that grazes the surface of the Sun, with $M=M_\odot$ and $b=R_\odot $, GR predicts $\delta_\tx{GR}=1.7516687^{\prime \prime}$. The observed value of light deflection on the solar surface is given by $\theta = \frac{1}{2}(1+\gamma)1.7516687^{\prime \prime}$ with $\gamma = 0.99992 \pm 0.00012$ \cite{Lambert2011}. Finally, the constraints on $\ell$ is given by $-1.1\times 10^{-10} \leq \ell \leq 5.4 \times 10^{-10}$.

\subsection{Time delay of light}

The Shapiro time-delay effect is observed by measuring the travel time of a light ray between two stations located at large distances from a massive object, such as the Sun. To simplify the analysis, we consider the motion of light on the equatorial plane ($\theta=\pi/2$) of a central celestial object. Since the light is moving along a null geodesic, we have $ds^{2}=0$, i.e.,
\beq
-A(r)dt^{2}+A(r)^{-1}dr^{2}+r^{2}d\phi^{2}=0.
\label{Null_geodesic}
\eeq
After applying the solution in Eq.~\eqref{Solution_u} and performing some simple algebra, we obtain the relation
\beq
r^{2}d\phi^{2}=\frac{b^2}{(1-\ell)\left(r^2-b^2\right)A(r)^2}dr^{2}.
\eeq
Substituting it into Eq.~\eqref{Null_geodesic} and considering that the Lorentz-violating parameter $\ell$ and $M/r$ (represented as $\frac{2GM}{c^2r}$ in the International System of Units) are both small quantities, we have
\beq
dt\approx \pm \fc{1-\ell}{\sqrt{r^2-b^2}} \lt(1+\fc{2 M}{r}+\fc{b^2 M}{r^3}\rt) rdr.
\eeq 
where the plus (minus) sign refers to the outgoing (infalling) light ray. 

Considering a light ray (or radar signal) that travels from an emitter situated at $r_\tx{E}$ to a receiver at $r_\tx{R}$, the time taken for the travel can be calculated as
\beqn
T\!&\!=\!&\! \lt(1-\ell\rt)\Bigg[\lt(\sqrt{r_\tx{E}^2-b^2}+\sqrt{r_\tx{R}^2-b^2}\rt)\nn\\
\!&\!+\!&\! M\lt(\frac{\sqrt{r_\tx{R}^2-b^2}}{r_\tx{R}}+\frac{\sqrt{r_\tx{E}^2-b^2}}{r_\tx{E}}\rt)\nn\\
\!&\!+\!&\! 2 M \lt( \log \left(\frac{\sqrt{r_\tx{E}^2-b^2}+r_\tx{E}}{b}\right)\rt.\nn\\
\!&+\!&\! \lt.\log \left(\frac{\sqrt{r_\tx{R}^2-b^2}+r_\tx{R}}{b}\right)\rt)\Bigg].
\eeqn
It is evident that when the Lorentz-violating parameter $\ell$ is absent, this result reduces to the prediction of GR. The first term inside the square bracket represents the time required for the light to propagate along a straight line from $r_\tx{E}$ to $r_\tx{R}$ in a flat spacetime, while the remaining terms represent Shapiro gravitational time delay. 

As the most precise test of time delay currently comes from the data obtained by the Cassini mission during its journey to Saturn \cite{Bertotti2003}, we consider the scenario where a radar signal is emitted from Earth at $r_\tx{E}=r_\oplus$, then grazes near the Sun and travels to the spacecraft located at $r_\tx{R}$ before returning to Earth along the same path. Then, due to the impact parameter $b\ll r_\oplus,r_\tx{R}$, the total round-trip time delay $T_\tx{tol}=2T$ can be approximated as 
\beqn
T_\tx{tol} \!&\!\approx\!&\! \lt(1-\ell\rt)\lt[2\lt(\sqrt{r_\oplus^2-b^2}+\sqrt{r_\tx{R}^2-b^2}\rt)\rt.\nn\\
\!&\!+\!&\! \lt. 4 M_\odot \lt(\log \lt(\frac{4 r_\oplus r_\tx{R}}{b^2}\rt)+1\rt)\rt] \nn\\
\!&\! \equiv \!&\! \lt(1-\ell\rt)\lt(T_0+ \delta T_\tx{GR} \rt),
\label{Time_delay}
\eeqn
where $T_0$ represents the round-trip time in flat spacetime and $\delta T_\tx{GR}$ represents the round-trip time delay caused by the curved spacetime. By noticing that $\ell T_0 \ll T_0$ and $\delta T_\tx{GR}\ll T_0$, the total excess time delay is given by
\beq
\delta T_\tx{tol}=T_\tx{tol} - T_0\approx \delta T_\tx{GR} - l T_0.
\eeq
Therefore, the time delay resulting from Lorentz violation is $\delta T_\tx{LV}=- l T_0$.

For the 2002 superior conjunction of Cassini, the spacecraft was at $r_\tx{R}\approx 8.43$AU from the Sun and the radar grazes near the Sun with an impact parameter $b\approx 1.6R_\odot$. Using Eq.~\eqref{Time_delay}, we obtain $T_0=9411.2$s and $\delta T_\tx{GR}=2.8\times 10^{-4}$s. The measured value of the post-Newtonian parameter $\gamma$ in the formula $\delta T=2(1+\gamma) M_\odot\log \lt(\frac{4 r_\oplus r_\tx{R}}{b^2}\rt)$ is reported as $\gamma=1+\lt(2.1 \pm 2.3 \rt)\times10^{-5}$ \cite{Bertotti2003}. Therefore, we obtain the constraints on the Lorentz-violating parameter $\ell$ as $-6.1\times10^{-13}\leq \ell\leq 2.8\times10^{-14}$.

\section{Conclusions}\label{Conclusions}

The static and spherically symmetric spacetime was investigated in the context of a gravity theory featuring LSB induced by a nonzero VEV of the KR filed. By considering both the presence and absence of the cosmological constant, we derived the exact solutions that exhibit new properties resulting from the Lorentz-violating effect. Furthermore, the thermodynamic properties of the AdS-Schwarzschild-like black hole were examined, revealing that the standard first law of thermodynamics and Smarr formula hold as usual. Interestingly, it was observed that the Lorentz-violating parameter $\ell$ does not impact the critical size, which determines the local and global stability of the black holes.

\begin{table}[h]
\begin{tabular}{|c|c|}
  \hline
Solar tests   & Constraints  \\
  \hline
Mercury precession& $-3.7 \times 10^{-12}\leq \ell \leq 1.9\times10^{-11}$\\
\hline 
Light deflection &$-1.1\times 10^{-10} \leq \ell \leq 5.4 \times 10^{-10}$\\
  \hline
Shapiro time delay &$-6.1\times10^{-13}\leq \ell\leq 2.8\times10^{-14}$\\
  \hline
\end{tabular}
\caption{Constraints on the Lorentz-violating parameter $\ell$ from Solar System tests.}
\label{Solar_tests}
\end{table}

In order to investigate the physical implications of the obtained solutions, we analyzed several classical gravitational experiments within the Solar System, including the perihelion precession of Mercury, deflection of light, and Shapiro time delay. Our analysis revealed that the Lorentz-violating effect does contribute to the corrections observed in these experiments. By utilizing the measured results from these experiments, we were able to constrain the value of the Lorentz-violating parameter $\ell$. The resulting constraints are listed in Tab.~\ref{Solar_tests}. It shows that the Shapiro time delay imposes the most stringent constraints on the Lorentz-violating effect, with the range of $-6.1\times10^{-13}\leq \ell\leq 2.8\times10^{-14}$. 

The Lorentz violation effect is also heavily constrained through Solar System tests in the bumblebee gravity model. Specifically, the perihelion precession of Mercury provides an upper bound of Lorentz-violating parameter $\ell<1.1\times10^{-11}$, the bending of light gives an upper bound of $\ell < 3.2\times 10^{-10}$, and the Shapiro time-delay effect observed during the Cassini mission establishes an upper bound of $\ell< 6.2 \times 10^{-13}$ \cite{Casana2018}. These constraints exhibit magnitudes similar to the limits presented in Tab.~\ref{Solar_tests}, indicating that the possible Lorentz-violating effect is highly restricted in nature.

\section*{ACKNOWLEDGMENTS}

We thank Yu-Xiao Liu, Shao-Wen Wei, Xiao-Long Du, Peng Cheng and Si-Jiang Yang for helpful discussions. Ke Yang also acknowledges the generous hospitality during a visit to the Lanzhou Center for Theoretical Physics,  where part of this work was completed. This work was supported by the National Natural Science Foundation of China under Grant No.~12005174.


\end{document}